\documentclass[onecolumn,preprintnumbers,amsmath,amssymb]{revtex4}

\usepackage{graphicx}
\usepackage{dcolumn}
\usepackage{bm}
\usepackage[utf8x]{inputenc}

\begin{document}

\title{S-wave Heavy Quarkonium Spectra: Mass, Decays and Transitions}

\author{Halil Mutuk}
 \email{halilmutuk@gmail.com}
\affiliation{Physics Department, Faculty of Arts and Sciences, Ondokuz Mayis University, 55139, Samsun, Turkey
}%


\begin{abstract}
In this paper we revisited phenomenological potentials. We studied S-wave heavy quarkonium spectra by two potential models. The first one is power potential and the second one is logarithmic potential. We calculated spin averaged masses, hyperfine splittings, decay constants, leptonic decay widhts, two-photon and two-gluon decay widths and some allowed  M1 transitions. We studied ground and 4 radially excited S-wave charmonium and bottomonium states via solving  nonrelativistic Schrödinger equation.  Although the potentials were studied in this paper is not directly QCD motivated potential, obtained results are agree well with experimental data and other theoretical studies. 
\end{abstract}

\maketitle

\section{\label{sec:level1}Introduction}
Quarkonium is the bound state of a quark and antiquark by the strong interaction. Excluding top quark $t$, the rest of the quarks $b,c,u,d,s$ can make quarkonium states. Large masses of $b$ and $c$ quarks comparing to others deserve additional interest since experimentally $b\bar{b}$ and $c\bar{c}$ systems give fruitful data in the experiments.  

Study of mesons composed of a heavy quark and antiquark gives a good basis for understanding of QCD dynamics. Discoveries of $c\bar{c}, b\bar{b}, c\bar{b}$ and $s\bar{b}$ systems provides quantitative tests for high energy physics. Heavy quarkonium spectra is very rich in its own basis due to presenting some puzzles. In the last decade experimental data of Belle, BaBar, BESIII, LHCb, CDF and other collaborations came up with interesting results for hadron physics. Emerging of so-called exotic states presented some puzzles for hadron spectroscopy. These states cannot be identified easily by the conventional quark model, i.e., they are not made of quark-antiquark and three quarks (or antiquark) scheme. The milestone of this exotic states was $X(3872)$. It was first observed by Belle Collaboration \cite{1}, then was confirmed by D0 \cite{2}, CDF II \cite{3} and BaBar \cite{4} collaborations. 

Threshold physics for quarkonium systems are very important to test QCD and its nature. Typically the threshold for $c\bar{c}$ system is roughly 3.71 GeV and for $b\bar{b}$ system is 10.50 GeV. The physics is not so much clean but at least more clearer for below than the above of threshold.  The states below the threshold are generally narrow and decay to open-flavour (non-zero flavour number) hadrons via electromagnetic and strong interactions. For above the open-flavour threshold, states are not much narrow and decays are dominated by strong interactions.  For example the so called $XYZ$ mesons all lie above the open charm threshold and physical observables such as masses have some puzzles \cite{5}.

Quarkonium spectrum can be studied by two approaches. The first one is stems from or better saying has an analogy of quantum mechanical description for energy levels of an atom. In this approach a potential is used to describe the formation of bound state. For light quarks with including relativistic effects  and heavy quarks below the open flavour threshold, this approach gives reliable results. But above the open flavour threshold bound state formation is complicated in view of potential model as mentioned before. The second approach is to use the principles of QCD. 

QCD is thought to be the {\it true} theory of strong interactions. The reason for that is  the coupling constant of QCD is of the order $1$ in lower energies, hence the truncation of the perturbative expansion can not be carried out. This coupling constant can be viewed as a perturbation parameter. So that some non-perturbative methods emerge such as QCD Sum Rules, lattice QCD and Effective Field Theories. These methods in general don't need the shape of potential, i.e. hadronization occurs without taking care of potential. In the case of heavy quarkonium systems, masses of constituents quarks are much bigger than the QCD hadronization  scale, $m_q >> \Lambda_{QCD}, ~ \Lambda_{QCD} \approx 200 $ MeV, so that they can be studied via nonrelativistic Schr\"{o}dinger equation with a phenomenological potential between the quark and antiquark. The fundamental assumption of constituent quark model is that the quarks are non-relativistic objects inside the hadron. For this reason, the total mass of a hadron is approximately equal to the sum of the masses of the quarks that compose hadron. The so called 'constituent quark masses' of quarks can be described as effective masses that  are chosen to fit the experimental results \cite{32}.
In constituent quark models and potential models quark masses are treated to be as free parameters and they are determined by fitting to experimental results. Constituent quark masses are generally larger than their pole masses. 

The nature of quarkonium potentials are somehow not clear. $q\bar{q}$ potentials cannot be derived from first principles of QCD. Therefore there is no dominant potential which interpret quarkonium spectrum exactly. There are nonrelativistic quark models \cite{6,7} and relativistic models \cite{8,9,10} to study spectrum. A general review can be found in \cite{11} and \cite{12} and references therein. 

There are a couple of methods to study hadron spectrum. Among QCD Sum Rules and lattice QCD, quark model has an advantage of studying excited states whereas QCD Sum Rules and lattice QCD focus on ground states. QCD has two distinct features: {\it asymptotic freedom} and {\it confinement}.  Hadronization occurs at large distances (lower momenta) where nonperturbative methods work but at very short distances (higher momenta) in which for example decay occurs, perturbative theory works. A potential must maintain this delicate balance between these two separate region in order to produce spectrum. This can be accomplished by assuming that interactions between quarks form a two body problem and they are heavy enough so that Schr\"{o}dinger equation can be used. The consequences of these two QCD features can be approximately modelled by Coulomb plus a linear potential, $V(r)=-\frac{1}{r}+r$.  In Coulomb part one-gluon-exchange is appropriate to consider physical phenomena whereas there is no consensus about the form of linear part. 

The other nonperturbative method is Effective Field Theory (EFT). In EFT focus on energy scales. Different scales are factorized which enables to describe physical phenomena in a specific range in terms of degrees of freedom. One have a {\it potential} which encodes the effect of degrees of freedom that can be extracted from QCD. These potentials can be classified as phenomenological (non-QCD like) and QCD motivated (or QCD inpired) potentials. For example,  Logarithmic \cite{35}, Richardson \cite{36},  Buchmüller-Tye \cite{37} potentials are phenomenological spin-independent potentials and the potential in \cite{38} is spin-dependent and velocity-dependent. 

There are many potential models to study heavy quarkonium spectra. In this work we used a power law potential \cite{13} and a logarithmic potential \cite{14} to study s-wave heavy quarkonium. We generated s-wave charmonium and bottomonium mass spectrum with the decays and M1 transitions.  At section 2 we give out theoretical model. In sections 3 and 4, we generate s-wave heavy quarkonium spectrum, decays and transitions. In last section we discuss our results. 

\section{\label{sec:level12}Formulation of Model}
 
The Hamiltonian  we consider is
\begin{equation}
H=M+\frac{p^2}{2\mu}+V\left( r \right)
\end{equation}
where $M=m_q+m_{\bar{q}}$,  $p$ is the relative momentum, $\mu$ is the reduced mass and $V\left( r \right)$
is the potential between quarks. The spectrum can be obtained via solving Schr\"{o}dinger equation
\begin{equation}
H\vert \Psi_n \rangle =E_n \vert \Psi_n \rangle
\end{equation}
with the harmonic oscillator wave function defined as

\begin{equation}
\Psi_{nlm}\left( r, \theta, \phi \right)=R_{nl}(r) Y_{lm}(\theta, \phi) \label{eq:psi}.
\end{equation}

Here $R_{nl}$ is the radial wave function given as

\begin{equation}
R_{nl}=N_{nl} r^l e^{-\nu r^2} L^{l+\frac{1}{2}}_{ \frac{n-l}{2}} \left( 2\nu r^2 \right) \label{eq:psir}
\end{equation}
with the associated Laguerre polynomials $L^{l+\frac{1}{2}}_{ \frac{n-l}{2}}$ and the normalization constant

\begin{equation}
N_{nl}=\sqrt{\sqrt{ \frac{2\nu^3}{\pi}} \frac{2 (\frac{n-l}{2})!\nu^l}{(\frac{n+l}{2}+1)!!}}.
\end{equation}
$Y_{lm}(\theta, \phi)$ is the well known  spherical harmonics. 
 
Armed with these one can obtain spin averaged spectra with the given potential. To solve Schrödinger equation, we used variational method. The procedure for this method is 
\begin{eqnarray}
E=\frac{\langle \Psi \vert H \vert \Psi \rangle}{\langle \Psi \vert \Psi \rangle }.
\end{eqnarray}

In Eqn. (\ref{eq:psir}), $\nu$ is treated as a variational parameter and it is determined for each state by minimizing the expectation value of the Hamiltonian.

In the following sections we study power law and logarithmic potentials  in order to obtain full spectrum.

\section{Mass Spectra of Power Law and Logarithmic Potentials}
Power law potential is given by \cite{15}
\begin{equation}
V\left( r \right)=-8.064 ~ \text{GeV} + 6.898 ~ \text{GeV} ~ r^{0.1}.
\end{equation}
The small power of $r$ refer to a situation in which the spacing of energy levels is independent of the quark masses. This situation is also valid for the purely logarithmic potential \cite{16}
\begin{equation}
V\left( r \right)=-0.6635  ~ \text{GeV} + 0.733 ~ \text{GeV}  ~ \text {Ln}  ~ (r \times 1  ~ \text{GeV} ).
\end{equation}

At first step we obtained spin averaged mass spectrum for $c\bar{c}$ and $b\bar{b}$ systems, respectively. The constituent quark masses are $m_c= 1.8 ~ \text{GeV} $ and $m_b= 5.174  ~ \text{GeV}$ for power-law potential and $m_c=1.5 ~ \text{GeV} $ and $m_b= 4.906 ~ \text{GeV}$ for the logarithmic potential. Table \ref{tab:table1} shows the charmonium spectrum and Table \ref{tab:table2} shows for the bottomonium.

\begin{table}[h]
\caption{\label{tab:table1}Spin-averaged mass spectrum of Charmonium (in MeV)}
\begin{ruledtabular}
\begin{tabular}{cccccc}
 State & Power & Logarithmic &\cite{17}&
 \cite{18}&\\
\hline
1S& 3067 & 3067 & 3067 & 3117 &\\
2S& 3701 & 3655 & 3667 & 3684 &\\
3S& 4054 & 3980 & 4121 & 4078 & \\
4S& 4306 & 4204 & 4513 & 4407 & \\
5S& 4504 & 4376 & 4866 &  &
\end{tabular}
\end{ruledtabular}

\end{table}

\begin{table}[h]
\caption{\label{tab:table2}Spin-averaged mass spectrum of Bottomonium (in MeV)}
\begin{ruledtabular}
\begin{tabular}{cccccc}
 State & Power & Logarithmic & \cite{17}&
 \cite{18}&\\
\hline
1S& 9473 & 9444 & 9443 & 9523 &\\
2S& 10049 & 10033 & 9729 & 10035 &\\
3S& 10384 & 10357 & 10312 & 10373 & \\
4S& 10624 & 10581 & 10593 &  & \\
5S& 10813 & 10753 & 10840 &  &
\end{tabular}
\end{ruledtabular}

\end{table}

Many potential models used to describe quarkonium states tend to be similar. Generally they have a Coulomb like term and a linear term. Maybe the most known is Cornell potential, $V(r)=-\frac{a}{r}+br$ where $a$ and $b$ are some parameters. In this potential first term is responsible for one-gluon exchange and the second term is for confinement. The other effects like relativistic or spin dependent interactions can be added to potential but most of them are considerably small compared to Coulomb or linear terms. The potentials we consider here doesn't have spin dependent terms. A general potential usually include spin-spin interaction, spin-orbit interaction and tensor force terms. To obtain whole picture, it is necessary to consider spin dependent terms within the potential. In this work we only take care of spin-spin interaction, i.e. hyperfine splitting. 

\subsection{Spin-spin Interaction}

Previous discussion of mass spectrum doesn't include spin dependent effects so it fails to interpret spin splittings. For example the mass splitting of $\eta_c (1S)$ and $J/\psi$ is $\Delta m=114 ~ MeV$ although they are the grounds states in $l=0$. Furthermore $c\bar{c}$ quarks have $s=0$ in $\eta_c (1S)$ whereas they have $s=1$ in $J/\psi$. The reason for this mass difference is allegedly to be spin dependent interactions. Generically, these spin dependent interactions can be obtained by considering nonrelativistic limit of Dirac equation. 

In the spectroscopic language, $q\bar{q}$ states are represented by  $J^{PC}$ or atomic notation $n^{2s+1}L_j$. $n$ is the principal quantum number, $s$ is the total spin of quarkonium system, $l$ is orbital angular momentum, $j$ is the total angular momenta, $P$ is the parity quantum number and $C$ is the charge conjugate quantum number. For mesons, parity number is given by $P=(-1)^{L+1}$ and charge number is given by $C=(-1)^{L+S}$ where $S$ is the total spin. 

Quark potential model gives a good description of the spin-averaged mass spectrum of hadrons since they are composite particles made up from quarks \cite{33}. Mass splitting is closely connected with the Lorentz-structure of the quark potential \cite{34}. The origin of the spin-spin interaction term lies in the one-gluon exchange term which is related to $1/r$. Spin is proportional of the magnetic moment of a particle. Magnetic moments generate short range fields $ \sim 1/r^3$. In the case of heavy quarkonium systems which are non-relativistic, wave functions of two particles overlap in a significant amount. This means that particles are very close to each other. So spin-spin interactions play a significant role in the dynamics.

The {\it spin-spin} interaction term of two particle can be written as
\begin{equation}
V_{SS}(r)=\frac{32\pi \alpha_s}{9 m_q m_{\bar{q}}}\vec{S}_q \cdot \vec{S}_{\bar{q}} \delta (\vec{r}).
\end{equation}
This term can explain $S$ wave splittings and has no contribution to $l \neq 0$ states. Putting this term into Schr\"{o}dinger equation we get

\begin{equation}
E_{HF}=\frac{32\pi \alpha_s}{9 m_q m_{\bar{q}}} \int d^3r \Psi^{\star}(\vec{r}) \Psi(\vec{r})  \delta(\vec{r}) \langle\vec{S}_q \cdot \vec{S}_{\bar{q}} \rangle.
\end{equation} 

Implementing Dirac-delta function property 
\begin{equation*}
\int f(x) \delta(x) dx= f(0),
\end{equation*}

we get

\begin{equation}
E_{HF}=\frac{32\pi \alpha_s}{9 m_q m_{\bar{q}}} \vert \Psi(0) \vert^2  \langle\vec{S}_q \cdot \vec{S}_{\bar{q}} \rangle.
\end{equation}

The matrix element of spin products can be obtained via 

\begin{eqnarray*}
\textbf{S}_1 \cdot \textbf{S}_2&=&\frac{1}{2} \left(\textbf{S}^2- \textbf{S}_1^2 - \textbf{S}_2^2 \right) \nonumber \\ &=& \frac{1}{2} \left(S(S+1)-\frac{3}{2} \right)
\end{eqnarray*}

so that

\begin{equation}
\langle\vec{S}_q \cdot \vec{S}_{\bar{q}} \rangle=\begin{cases}
    \frac{1}{4}, ~ \text{for $\vec{S}=1$}\\
   -\frac{3}{4}, ~ \text{for $\vec{S}=0$}.
  \end{cases}
\end{equation}

Therefore we obtain hyperfine splittings energy as

\begin{equation}
 E_{HF}=\begin{cases}
    \frac{8 \pi \alpha_s}{9 m_q m_{\bar{q}}}\vert \Psi(0) \vert^2, & \text{for $\vec{S}=1$}\\
   -\frac{8 \pi \alpha_s}{3 m_q m_{\bar{q}}}\vert \Psi(0) \vert^2, & \text{for $\vec{S}=0$}.
  \end{cases}
\end{equation}

Here $\Psi(0) $ is the wave function at the origin and can be obtained by the following relation
\begin{equation}
\vert \Psi(0) \vert^2=\frac{\mu}{2\pi \bar{h}} \left\langle \frac{d V(r)}{dr}\right\rangle.
\end{equation}
\\
Expectation value is obtained by the wave function given in (\ref{eq:psi}). S-wave charmonium and bottomonium masses can be seen in Tables \ref{tab:table3} and \ref{tab:table4}. In this calculation, $\alpha_s$ is taken to be 0.37 for charmonium and 0.26 for bottomonium \cite{17}.\\

\begin{table}[h]
\caption{\label{tab:table3}Charmonium mass spectrum (in MeV). In ref \citep{22} LP denotes linear potential and SP denotes screened potential. }
\begin{ruledtabular}
\begin{tabular}{cccccccccc}
 State &Exp. \cite{19}& Power & Logarithmic & \cite{20}&
 \cite{21}& \cite{22} LP & \cite{22} SP \\
\hline
$\eta_c$(1S)& 2984 & 2980 & 2954 & 2979 & 2982 & 2983 & 2984\\
$\eta_c$(2S)& 3639 & 3624 & 3555 & 3623 & 3630 & 3635 & 3637\\
$\eta_c$(3S)&      & 3983 & 3887 & 3991 & 4043 & 4048 & 4004\\
$\eta_c$(4S)&      & 4240 & 4117 & 4250 & 4384 & 4388 & 4264\\
$\eta_c$(5S)&      & 4441 & 4294 & 4446 &      & 4690 & 4459\\ 
$J/\psi$    & 3097 & 3096 & 3104 & 3097 & 3090 & 3097 & 3097\\
$\psi$(2S)  & 3686 & 3727 & 3689 & 3673 & 3672 & 3679 & 3679\\
$\psi$(3S)  & 4040 & 4078 & 4011 & 4022 & 4072 & 4078 & 4030\\
$\psi$(4S)  &      & 4328 & 4233 & 4273 & 4406 & 4412 & 4281\\
$\psi$(5S)  &      & 4525 & 4403 & 4463 &      & 4711 & 4472\\
\end{tabular}
\end{ruledtabular}

\end{table}

\begin{table}[h]
\caption{\label{tab:table4}Bottomonium  mass spectrum (in MeV). }
\begin{ruledtabular}
\begin{tabular}{cccccccccc}
 State &Exp. \cite{23}& Power & Logarithmic & \cite{24}&
 \cite{25}& \cite{26}  & \cite{27}  \\
\hline
$\eta_b$(1S)  & 9399  & 9452  & 9420  & 9389  & 9390 & 9402 & 9455 \\
$\eta_b$(2S)  & 9999  & 10030 & 10011 & 9987  & 9990 & 9976 & 9990 \\
$\eta_b$(3S)  &       & 10367 & 10338 & 10330 & 10326 & 10336 & 10330 \\
$\eta_b$(4S)  &       & 10608 & 10562 & 10595 & 10584 & 10623 &  \\
$\eta_b$(5S)  &       & 10798 & 10735 & 10817 & 10800 & 10869 &  \\ 
$\Upsilon$(1S)& 9460  & 9480  & 9452  & 9460  & 9460 & 9465 & 9502 \\
$\Upsilon$(2S)& 10023 & 10055 & 10040 & 10016 & 10015 & 10003 & 10015 \\
$\Upsilon$(3S)& 10355 & 10393 & 10364 & 10351 & 10343 & 10354 & 10349 \\
$\Upsilon$(4S)& 10579 & 10629 & 10588 & 10611 & 10597 & 10635 & 10607 \\
$\Upsilon$(5S)& 10865 & 10818 & 10759 & 10831 & 10811 & 10878 & 10818 \\
\end{tabular}
\end{ruledtabular}

\end{table}

As can be seen from Tables \ref{tab:table3} and \ref{tab:table4}, our results are compatible with both experimental and theoretical results. 

\section{Dynamical Properties}
\subsection{Decay constants}
Leptonic decay constants give information about short distance structure of hadrons. In the experiments this regime is testable since the momentum transfer is very large. The pseudoscalar ($f_p$) and the vector ($f_v$) decay constants are defined respectively through the matrix elements \cite{18}

\begin{equation}
p^{\mu}f_p=i \langle 0  \vert \bar{\Psi} \gamma^{\mu} \gamma^{5} \Psi \vert p \rangle
\end{equation}
and
\begin{equation}
m_v f_v \epsilon^{\mu}= \langle 0  \vert \bar{\Psi} \gamma^{\mu}  \Psi \vert v \rangle.
\end{equation}

In the first relation, $p^{\mu}$ is meson momentum and $\vert p \rangle$ is pseudoscalar meson state. In the second relation, $m_v$ is mass, $\epsilon^{\mu}$ is the polarization vector and $\vert v \rangle$ is the state vector of meson. 

The matrix elements can be calculated by quark model wave function in the momentum space. The result is 
\begin{eqnarray}
f_p &=&\sqrt{\frac{3}{m_p}} \int \frac{d^3k}{(2\pi)^3} \sqrt{1+\frac{m_q}{E_k}} \sqrt{1+\frac{m_{\bar{q}}}{E_{\bar{k}}}} 
 \left( 1-\frac{k^2}{(E_k+m_q)(E_{\bar{k}}+m_{\bar{q}})} \right) \phi (\vec{k})
\end{eqnarray}
for pseudoscalar meson and

\begin{eqnarray}
f_v &=&\sqrt{\frac{3}{m_v}} \int \frac{d^3k}{(2\pi)^3} \sqrt{1+\frac{m_q}{E_k}} \sqrt{1+\frac{m_{\bar{q}}}{E_{\bar{k}}}} 
  \left( 1+\frac{k^2}{3(E_k+m_q)(E_{\bar{k}}+m_{\bar{q}})} \right) \phi (\vec{k})
\end{eqnarray}
for the vector meson \cite{18}. 

In the nonrelativistic limit, these two equations take a simple form which is known to be Van Royen and Weisskopf relation \cite{28} for the meson decay constants

\begin{equation}
f^2_{p/v}=\frac{12 \vert \Psi_{p/v}(0) \vert^2}{m_{p/v}}.
\end{equation}

The first order correction which is also known as QCD correction factor is given by

\begin{equation}
\bar{f}_{p/v}^2=\frac{12 \vert \Psi_{p/v}(0) \vert^2}{m_{p/v}} C^2(\alpha_s)
\end{equation}
where $C(\alpha_s)$ is given by \cite{29}

\begin{equation}
C(\alpha_s)=1-\frac{\alpha_s}{\pi} \left(\Delta_{p/v}-\frac{m_q-m_{\bar{q}}}{m_q+m_{\bar{q}}} \text{ln}  \frac{m_q}{m_{\bar{q}}} \right)
\end{equation}
 
where $\Delta_p=2$ for pseudoscalar mesons and  $\Delta_v=8/3$ for vector mesons. Decay constants are given in Tables \ref{tab:table5} 

 \begin{table}[h]
\caption{\label{tab:table5}Pseudoscalar decay constants (in MeV).}
\begin{ruledtabular}
\begin{tabular}{ccccccccccc}
 State &Exp. \cite{23}& Power $f_p$  & Power  $\bar{f_p}$& Logarithmic  $f_p$ & Logarithmic  $\bar{f_p}$ & \cite{17}  $f_p$& \cite{17} $\bar{f_p}$  &
 \cite{18}  \\
\hline
$\eta_c$(1S)& 335 $\pm$ 75 & 543 & 415 & 578 & 442 & 471& 360&402\\
$\eta_c$(2S)&  & 473 & 362 & 497 & 380 & 374 &286 &240\\
$\eta_c$(3S)&      & 330 & 252 & 442 & 338 &332& 254&193\\
$\eta_c$(4S)&      & 325 & 248 & 412 & 315 &312& 239 \\
$\eta_c$(5S)&      & 253 & 193 & 387 & 304 & \\ 
$\eta_b$(1S)&   & 517  & 431  & 585  & 488 &834 &694 & 599 \\
$\eta_b$(2S)&   & 479 & 400 & 535  & 447 &567 &472 & 411\\
$\eta_b$(3S)&       & 345 & 288 & 504 & 421 &508& 422& 354\\
$\eta_b$(4S)&       & 313 & 261 & 482 & 402 &481 &401\\
$\eta_b$(5S)&       & 283 & 236 & 465 & 388 &  \\ 
 \\
\end{tabular}
\end{ruledtabular}

\end{table}

 and \ref{tab:table6} 
 
  \begin{table}[h]
\caption{\label{tab:table6}Vector decay constants (in MeV).}
\begin{ruledtabular}
\begin{tabular}{ccccccccccc}
 State &Exp. \cite{23}& Power $f_v$  & Power $\bar{f_v}$& Logarithmic $f_v$ & Logarithmic $\bar{f_v}$ & \cite{17}  $f_p$&  \cite{17} $\bar{f_p}$&
 \cite{18}  \\
\hline
$J/\psi$& 335 $\pm$ 75 & 529 & 363 & 563 & 386 & 462& 317&393\\
$\psi$(2S)& 279 $\pm$ 8  & 463 & 318 & 487 & 334 & 369& 253 & 293\\
$\psi$(3S)& 174 $\pm$ 18 & 324 & 222 & 436 & 299 &329 &226 & 258\\
$\psi$(4S)&      & 319 & 219 & 406 & 279 & 310 &212 \\
$\psi$(5S)&      & 248 & 170 & 382 & 262 & 290 &199\\ 
$\Upsilon$(1S)& 708 $\pm$ 8  & 516  & 402  & 584  & 455 &831& 645 & 665 \\
$\Upsilon$(2S)& 482 $\pm$ 10 & 482 & 373 & 535  & 416 &566 &439 & 475\\
$\Upsilon$(3S)& 346 $\pm$ 50 & 350 & 269 & 504 & 393 &507 &393& 418\\
$\Upsilon$(4S)& 325 $\pm$ 60 & 316 & 243 & 482 & 375 &481 &373&388\\
$\Upsilon$(5S)& 369 $\pm$ 93& 285 & 222 & 464 & 362 &458 &356 & 367  \\ 
 \\
\end{tabular}
\end{ruledtabular}

\end{table}
 
\subsection{Leptonic decay widths}
 
Leptonic decay of a vector meson with $J^{PC}=1^{--}$ quantum numbers can be pictured by the following annihilation via a virtual photon
\begin{equation}
V(q\bar{q}) \rightarrow \gamma \rightarrow e^+ e^-.
\end{equation}
 This state is neutral and in principle can decay into a different lepton pair rather than electron-positron pair. The above amplitude can be calculated by the Van Royen and Weisskopf relation \cite{11}
 
 \begin{equation}
 \Gamma (n^3 \text{S}_1 \rightarrow e^+ e^-)=\frac{16 \pi \alpha^2 e_q^2 \vert \Psi(0) \vert^2}{m_n^2} \times (1-\frac{16\alpha_s}{3\pi}+ \cdots)
 \end{equation}
 where $\alpha=\frac{1}{137}$ is the fine structure constant, $e_q$ is the quark charge, $m_n$ is the mass of $n^3 \text{S}_1$ state and $ \vert \Psi_{p/v}(0) \vert $ is the wave function at the origin of initial state. The term in the parenthesis is the first order QCD correction factor while $\cdots$ represents higher corrections. The obtained values for leptonic decay widths can be found in Tables \ref{tab:table7} and \ref{tab:table8} for charmonium and bottomonium, respectively. 
 
 \begin{table*}[h]
\caption{\label{tab:table7}Charmonium leptonic decay widths (in keV). The
widths calculated with and without QCD corrections are denoted by  $\Gamma_{l^+l^-}$ and $\Gamma^0_{l^+l^-}$.}
\begin{ruledtabular}
\begin{tabular}{cccccccccc}
 &\multicolumn{2}{c}{Power}&\multicolumn{2}{c}{Logarithmic}&\multicolumn{2}{c}{\cite{5}}&\multicolumn{2}{c}{\cite{17}}& Exp. \citep{23}\\
State &   $\Gamma^0_{l^+l^-}$   & $\Gamma_{l^+l^-}$ & $\Gamma^0_{l^+l^-}$  
&$\Gamma_{l^+l^-}$&  $\Gamma^0_{l^+l^-}$  & $\Gamma_{l^+l^-}$&  $\Gamma^0_{l^+l^-}$  & $\Gamma_{l^+l^-}$\\ \hline
 $J/\psi$& 3.435 & 1.277 & 3.154 & 1.173 & 11.8 & 6.60 & 6.847 & 2.536 & 5.55 $\pm$ 0.14 $\pm$ 0.02\\
 $\psi$(2S)& 2.880 & 1.071 &2.362 & 0.878 & 4.29 & 2.40 & 3.666 & 1.358 & 2.33 $\pm$ 0.07 \\
$\psi$(3S)&2.153 & 0.800 &1.888 & 0.702 & 2.53 & 1.42 & 2.597 & 0.962 & 0.86 $\pm$ 0.07 \\
$\psi$(4S)&1.839&0.684&1.642&0.610&1.73&0.97& 2.101&0.778& 0.58 $\pm$ 0.07\\
$\psi$(5S)&1.590 &0.591&1.551 &0.576&1.25&0.70& 1.710&0.633\\
\end{tabular}
\end{ruledtabular}
\end{table*}

  \begin{table*}[h]
\caption{\label{tab:table8}Bottomonium leptonic decay widths (in keV). The
widths calculated with and without QCD corrections are denoted by  $\Gamma_{l^+l^-}$ and  $\Gamma^0_{l^+l^-}$.}
\begin{ruledtabular}
\begin{tabular}{cccccccccc}
 &\multicolumn{2}{c}{Power}&\multicolumn{2}{c}{Logarithmic}&\multicolumn{2}{c}{\cite{24}}&\multicolumn{2}{c}{\cite{17}}& Exp. \citep{23}\\
State &  $\Gamma^0_{l^+l^-}$  & $\Gamma_{l^+l^-}$& $\Gamma^0_{l^+l^-}$
&$\Gamma_{l^+l^-}$&  $\Gamma^0_{l^+l^-}$  & $\Gamma_{l^+l^-}$&  $\Gamma^0_{l^+l^-}$  & $\Gamma_{l^+l^-}$\\ \hline
$\Upsilon$(1S)& 0.817 & 0.456 & 0.847 & 0.473 & 2.31 & 1.60 & 1.809 & 0.998 & 1.340 $\pm$ 0.018\\
$\Upsilon$(2S)& 0.686 & 0.383 & 0.709 & 0.396 & 0.92 & 0.64 & 0.797 & 0.439 & 0.612 $\pm$ 0.011 \\
$\Upsilon$(3S)& 0.610 & 0.340 & 0.630 & 0.352 &0.64 & 0.44 &  0.618 & 0.341 & 0.443 $\pm$ 0.008 \\
$\Upsilon$(4S)& 0.557 & 0.311 & 0.576 & 0.322 & 0.51 & 0.35 & 0.541 & 0.298 & 0.272 $\pm$ 0.029\\
$\Upsilon$(5S)& 0.526 & 0.294 & 0.535 &0.299&0.42&0.29& 0.481&0.265& 0.31 $\pm$ 0.07\\
\end{tabular}
\end{ruledtabular}
\end{table*}

\newpage 
\subsection{Two-photon decay width}
 
$^1\text{S}_0$ states with $J^{PC}=0^{-+}$ quantum number of charmonium and bottomonium can decay into two photons. In the nonrelativistic limit, the decay width for $^1\text{S}_0$ state can be written as \cite{30}

\begin{equation}
\Gamma(^1\text{S}_0 \rightarrow \gamma \gamma)=\frac{12 \pi \alpha^2 e_q^4 \vert \Psi(0) \vert^2 }{m_q^2}  \times (1-\frac{3.4\alpha_s}{\pi}).
\end{equation}
The term in the paranthesis is the first order QCD radiative correction. The results are listed in Table \ref{tab:table9}. 

\begin{table*}[h]
\caption{\label{tab:table9}Two-photon decay widths (in keV).  The
widths calculated with and without QCD corrections are denoted by  $\Gamma_{\gamma \gamma}$ and $\Gamma^0_{\gamma \gamma}$ }
\begin{ruledtabular}
\begin{tabular}{cccccccccc}
 &\multicolumn{2}{c}{Power}&\multicolumn{2}{c}{Logarithmic}& \multicolumn{2}{c} {\cite{17}}& \cite{8}& \cite{18} &Exp. \citep{23}\\
State &  $\Gamma^0_{\gamma \gamma}$  & $\Gamma_{\gamma \gamma}$& $\Gamma^0_{\gamma \gamma}$
&$\Gamma_{\gamma \gamma}$&  $\Gamma^0_{\gamma \gamma}$  & $\Gamma_{\gamma \gamma}$&  $\Gamma_{\gamma \gamma}$  & $\Gamma_{\gamma \gamma}$\\ \hline
$\eta_c$(1S)& 1.10 & 0.664 & 1.450 & 0.869 & 11.17 & 6.668 & 3.69 & 7.18& 7.2 $\pm$ 0.7 $\pm$ 0.2\\
$\eta_c$(2S)& 0.987 & 0.592 & 1.291 & 0.774 & 8.48 & 5.08 & 1.4 &1.71&   \\
$\eta_c$(3S)& 0.907 & 0.543 & 1.184 & 0.710 & 7.57 & 4.53 &  0.930 & 1.21 \\
$\eta_c$(4S)& 0.847 & 0.508 & 1.105 & 0.662 &  &  & 0.720 & &\\
$\eta_c$(5S)& 0.801 & 0.480 & 1.044 & 0.620 & && &&  \\
$\eta_b$(1S)& 0.277 & 0.199 & 0.277 & 0.199 & 0.58 & 0.42 & 0.214 & 0.45&    \\
$\eta_b$(2S)& 0.212 & 0.153 & 0.246 & 0.177 & 0.29 & 0.20 & 0.121 & 0.11& \\
$\eta_b$(3S)& 0.195 & 0.142 & 0.226 & 0.162 & 0.24 & 0.17 & 0.09 & 0.063 & \\
$\eta_b$(4S)& 0.188 & 0.136 & 0.211 & 0.151 &         &      & 0.07  &    \\
$\eta_b$(5S)& 0.176 & 0.129 & 0.199 & 0.143      &       &      &     &    \\
\end{tabular}
\end{ruledtabular}
\end{table*}

\subsection{Two-gluon decay width}

Two gluon decay width is given by \cite{30}
\begin{eqnarray}
\Gamma (^1\text{S}_0 \rightarrow gg)= \frac{8\pi \alpha_s^2  \vert \Psi(0) \vert^2 }{3m_q^2}  \times  \begin{cases}
    (1+4.8 \alpha_s \pi) ~ \text{for} ~ \eta_c  \\
  (1+4.4 \alpha_s \pi) ~ \text{for} ~   \eta_b.
  \end{cases}
\end{eqnarray}

The terms in the paranthesis refer to QCD corrections. The obtained results are given in Table \ref{tab:table10}.

\begin{table*}[h]
\caption{\label{tab:table10}Two-gluon decay widths (in MeV).  The
widths calculated with and without QCD corrections are denoted by  $\Gamma_{gg}$ and $\Gamma^0_{gg}$ }
\begin{ruledtabular}
\begin{tabular}{cccccccccc}
 &\multicolumn{2}{c}{Power}&\multicolumn{2}{c}{Logarithmic}& \multicolumn{2}{c} {\cite{17}}& \cite{31} &Exp. \citep{23}\\
State &  $\Gamma^0_{gg}$  & $\Gamma_{gg}$& $\Gamma^0_{gg}$
&$\Gamma_{gg}$&  $\Gamma^0_{gg}$  & $\Gamma_{gg}$&  $\Gamma^0_{gg}$  & $\Gamma_{gg}$\\ \hline
$\eta_c$(1S)& 32.04 & 50.15  & 41.93 & 32.44& 50.82 & 66.68 & 15.70 & 26.7 $\pm$ 3.0\\
$\eta_c$(2S)& 28.55 & 44.70 & 37.32 & 24.64 & 38.61 & 5.08 & 8.10 &14  $\pm$ 7 \\
$\eta_c$(3S)& 26.22 & 41.04 & 34.23 & 53.59 & 21.99 &  &   \\
$\eta_c$(4S)& 24.50 & 38.35 & 31.96 & 50.03 &  &  \\
$\eta_c$(5S)& 23.15 & 36.24 & 30.18 & 47.24 & &  \\
$\eta_b$(1S)& 5.50 & 7.50 & 12.82 & 17.49 & 13.72 & 18.80 & 11.49&    \\
$\eta_b$(2S)& 4.90 & 6.69 & 11.41 & 15.56 & 6.73 & 9.22 & 5.16 \\
$\eta_b$(3S)& 4.50 & 6.14 & 10.46 & 14.28 & 5.58 & 7.64 &3.80& \\
$\eta_b$(4S)& 4.20 & 5.74 & 9.77 & 13.33 &         &      &   &    \\
$\eta_b$(5S)& 3.97 & 5.42 & 9.22 & 12.58      &       &      &     &    \\
\end{tabular}
\end{ruledtabular}
\end{table*}

\subsection{M1 transitions}

M1 (magnetic dipole transition) decay widths can give more information about spin-singlet states. Moreover M1 transition rates show the  validity of theory against experiment \cite{21}. Magnetic transitions conserve both parity and orbital angular momentum of the initial and final states but in the M1 transitions the spin of the state changes. M1 width between two S wave states is given by \cite{6}
\begin{eqnarray}
\Gamma (n^3\text{S}_1 \rightarrow n^\prime ~ ^1\text{S}_0+ \gamma)&=& \frac{4 \alpha e_q^2 E_{\gamma}^3}{3m_q^2} \left( 2J_f+1 \right)\nonumber \\
& \times & \vert \langle f \vert j_0(kr/2) \vert i \rangle \vert^2,
\end{eqnarray}
where $E_{\gamma}=\frac{M_i^2-M_f^2}{2M_i}$ is the photon energy and $j_0(x)$ is the zeroth-order spherical Bessel function. In the case of small $E_{\gamma}$, spherical Bessel function $j_0(kr/2) $ tends to 1,  $j_0(kr/2)  \rightarrow 1$. Thus transitions between the same principal quantum numbers, $n^\prime =n$ are favored and usually known to be 'allowed'. In the other case, when $n^\prime \neq n$ the overlap  integral between initial and final state is 0 and generally designated as 'forbidden' transitions. The obtained transition rates for the 'allowed' ones of s-wave charmonium and bottomonium states are given in Tables \ref{tab:table11} and \ref{tab:table12}, respectively.

\begin{table*}[h]
\caption{\label{tab:table11}Radiative M1 decay widths of charmonium. In Ref. \cite{22} LP stands for linear potential and SP stands for screened potential. }
\begin{ruledtabular}
\begin{tabular}{cccccccccc}
Inıtial & Final& \multicolumn{2}{c}{Power}& \multicolumn{2}{c}{Logarithmic}&  \cite{17}& \multicolumn{2}{c}{\cite{22}}  &Exp. \cite{23}\\
&& $E_\gamma ~(\text{MeV})$  & $\Gamma ~(\text{keV})$  & $E_\gamma~(\text{MeV})$ & $\Gamma ~(\text{keV})$
&$\Gamma ~(\text{keV})$&  $\Gamma_{LP} ~(\text{keV})$  & $\Gamma_{SP} ~(\text{keV})$& $\Gamma ~(\text{keV})$ \\ \hline
$J/\psi$ & $\eta_c$(1S) & 114.9 & 1.96   & 113.8  & 2.83 & 3.28 & 2.39&  2.44  & 1.13 $\pm$ 0.35  \\
$\psi$(2S) & $\eta_c$(2S)& 111.5 & 1.39 & 101.5 & 2.01  & 1.45 & 0.19  & 0.19  &\\
$\psi$(3S) & $\eta_c$(3S)& 93.8 & 1.10 & 93.8 & 1.59  &  & 0.051 & 0.088  &\\
$\psi$(4S) & $\eta_c$(4S)& 87.1 & 0.88 & 87.1 & 1.27  &  &  &   &\\
$\psi$(5S) & $\eta_c$(5S)& 83.2 & 0.74 & 83.2 & 1.10  &  &  &   &\\

\end{tabular}
\end{ruledtabular}
\end{table*}
 
 \begin{table*}[h]
\caption{\label{tab:table12}Radiative M1 decay widths of bottomonium. }
\begin{ruledtabular}
\begin{tabular}{cccccccccc}
Inıtial & Final& \multicolumn{2}{c}{Power}& \multicolumn{2}{c}{Logarithmic}&  \cite{9}& \cite{26} & \cite{27}  &Exp. \cite{23}\\
&& $E_\gamma ~(\text{MeV})$  & $\Gamma ~(\text{eV})$  & $E_\gamma~(\text{MeV})$ & $\Gamma ~(\text{eV})$
&$\Gamma ~(\text{eV})$&  $\Gamma ~(\text{eV})$  & $\Gamma~(\text{eV})$& $\Gamma ~(\text{eV})$ \\ \hline
$\Upsilon$(1S) & $\eta_b$(1S) & 27.9 & 0.88   & 31.9  & 1.46 & 5.8 & 10&  9.34  &   \\
$\Upsilon$(2S) & $\eta_b$(2S)& 24.9 & 0.62 & 28.9 & 1.09  & 1.4 & 0.59  & 0.58  &\\
$\Upsilon$(3S) & $\eta_b$(3S)& 25.9 & 0.54 & 25.9 & 0.78  &  0.8 & 0.25 & 0.66  &\\
$\Upsilon$(4S) & $\eta_b$(4S)& 20.9 & 0.37 & 20.9 & 0.41  &  &  &   &\\
$\Upsilon$(5S) & $\eta_b$(5S)& 19.9 & 0.32 & 19.9 & 0.35  &  &  &   &\\

\end{tabular}
\end{ruledtabular}
\end{table*}
\newpage

\section{Discussion and Conclusion}
In this paper, we have computed spin averaged masses, hyperfine splittings, decay constants, leptonic decay widths, two photon and gluon decay widths and allowed M1 partial widths of S-wave heavy quarkonium states.  Nonrelativistic power and logarithmic potentials were used and compared with the experimental and theoretical studies.

Spin averaged mass give idea about the formulation of model since the results are close to experimental values since contributions from spin dependent interactions are small compared to contribution from potential part. If one ignore all spin dependent interactions, obtained results under this assumption is thought be averages over related spin states for principal quantum number.  Our results are in good agreement with pther studies.  

Hyperfine splittings is suggestive for Lorentz nature of confining heavy qaurkonium potentials. Including hyperfine splittings in the Hamiltonian, we obtained spectrum of both charmonium and bottomonium S-wave states. We calculated 5 states of heavy quarkonium by solving  nonrelativistic Schr\"{o}dinger equation with the harmonic oscillator wave function. The spectra for both power and logarithmic potentials agree well with experimental and other theoretical studies. The mass differences can be seen in \ref{tab:table13} for charmonium and \ref{tab:table14}, respectively. 

\begin{table}[h]
\caption{\label{tab:table13}Mass differences of S-wave charmonium states (in MeV). }
\begin{ruledtabular}
\begin{tabular}{cccccccccc}
 State &Exp. \cite{19}& Power & Logarithmic & \cite{20}&
 \cite{21}& \cite{22} LP & \cite{22} SP \\
\hline
$J/\psi$-$\eta_c$(1S)& 113 & 116 & 150 & 118 & 108 & 114 & 113\\
$\psi$(2S)-$\eta_c$(2S)& 47 & 103 & 134 & 50 & 42 & 44 & 42\\
$\psi$(3S)-$\eta_c$(3S)&      & 95 & 124 & 31 & 29 & 30 & 26\\
$\psi$(4S)-$\eta_c$(4S)&      & 88 & 116 & 23 & 22 & 24 & 17\\
$\psi$(5S) - $\eta_c$(5S)&      & 84 & 109 & 17 &      & 21 & 13\\ 
\end{tabular}
\end{ruledtabular}

\end{table}

\begin{table}[h]
\caption{\label{tab:table14}Mass differences of S-wave bottomonium states (in MeV). }
\begin{ruledtabular}
\begin{tabular}{cccccccccc}
 State &Exp. \cite{23}& Power & Log & \cite{24}&
 \cite{25}& \cite{26}  & \cite{27}  \\
\hline
$\Upsilon$(1S)-$\eta_b$(1S)  & 61 & 28  & 32  & 71  & 70 & 63 & 47 \\
$\Upsilon$(2S)-$\eta_b$(2S)  & 24  & 25 & 29 & 29  & 25 & 27 & 25 \\
$\Upsilon$(3S)-$\eta_b$(3S)  &       & 26 & 26 & 21 & 17 & 18 & 19 \\
$\Upsilon$(4S)-$\eta_b$(4S)  &       & 21 & 26 & 16 & 13 & 12 &  \\
$\Upsilon$(5S)-$\eta_b$(5S)  &       & 20 & 24 & 14 & 11 & 9 &  \\ 
\end{tabular}
\end{ruledtabular}

\end{table}

For charmonium case, mass difference for $J/\psi$-$\eta_c$(1S) agrees well with experimental data and theoretical results. For other states, i.e. radially excited states, power and logarithmic mass differences are bigger than the others. In case of bottomonium spectra, mass differences for all states are in good agreement with experimental results and theoretical studies.

Decay constants, of pseudoscalar and vector heavy quarkonium states are studied via Van Royen–Weisskopf relation. We got good results with experimental and theoretical studies within a few MeV. Including QCD corrrections, the results better agree in some states such as $J/\psi$ and $\eta_c$(1S). 

Leptonic decay widths of vector states are calculated and the results including QCD correction are comparable with experimental results. For some states such as $J/\psi$ and $\Upsilon$(1S), leptonic decay constants without QCD correction are more close to experimental results. This could be related to short range phenomena of potential models. 

Two-photon and two-gluon decay widths are also calculated within our formulation. For two-photon decay widhts, our results except $\eta_c$(1S) are comparable with other results. This situation is more better for ground and radially excites states in the case of two-gluon decay widhts. 

Obtained results for radiative M1 transitions of S-wave charmonium and bottomonium states are compatible with the other results.

To sum up, calculated results in this paper agree and quite agree well with experimental data and theoretical results. We hope that our work for S-wave heavy quarkonium system with power and logarithmic potential will be useful in the future studies. Different potential models can be studied also. We choose S-wave states since ground states and radially excited states are accessible from other models such as QCD Sum Rules. To test various predictions from current models, more data are necessary by forthcoming experiments.


\begin{thebibliography}{}
\bibitem{1}
S. K. Choi et al (Belle Collaboration), Observation of a Narrow Charmoniumlike State in Exclusive $B^{\pm}\rightarrow K^{\pm} \pi^+ \pi^- J/\psi $ Decays, Phy. Rev. Lett.\textbf{91}, 262001  (2003).

\bibitem{2}
V. M. Abazov  et al (D0 Collaboration), Observation and Properties of the $X(3872)$ Decaying to $J/\psi \pi^+ \pi^- $ in $p\bar{p}$ Collisions at $\sqrt{s}= \text{1.96} ~ \text{TeV}$, Phy. Rev. Lett. \textbf{93}, 162002 (2004).

\bibitem{3}
D. Acosta et al (CDF II Collaboration), Observation of the Narrow State $X(3872)\rightarrow J/\psi \pi^+ \pi^-  $ in $p\bar{p}$ Collisions at $\sqrt{s}= \text{1.96} ~ \text{TeV}$,  Phy. Rev. Lett. \textbf{93}, 072001  (2004).

\bibitem{4}
 Aubert B et al (BABAR Collaboration), Study of $B^- \rightarrow J/\psi K^- \pi^+ \pi^- $ decay and the measurement of the $B^- \rightarrow X(3872) K^- $ branching fraction, Phys. Rev. D \textbf{71}, 071103 (2005).
 
 \bibitem{5} 
  B. Q. Li,  K. T. Chao, Higher Charmonia and X,Y,Z states with Screened Potential,  Phys. Rev. D \textbf{79}, 094004 (2009).
 
\bibitem{6}
 E.  Eichten, K.  Gottfried,  T. Kinoshita, K. D. Lane and  T.-M. Yan, Charmonium: The Model,   Phys. Rev. D \textbf{17} 3090 (1978). 
 
\bibitem{7}
D. Gromes, On the effective quark potential in baryons,  Nucl. Phys. B \textbf{130}, 18 (1977). 

\bibitem{8}
 S. Godfrey and  N. Isgur, Mesons in a Relativized Quark Model with Chromodynamics,   Phys. Rev. D \textbf{32}, 189 (1985). 

\bibitem{9}
 D. Ebert, R. N. Faustov and  V. O. Galkin, Properties of heavy quarkonia and B-c mesons in the relativistic quark model,  Phys. Rev. D \textbf{67},  014027 (2003). 

\bibitem{10}
 S. Godfrey,  K. Moats and  E. S. Swanson, B and B-s Spectroscopy,  Phys. Rev. D \textbf{94} 054025 (2016). 
 
 \bibitem{11}
 E. Eichten, S. Godfrey, H. Mahlke, J. L. Rosner, Quarkonia and Their Transitions, Rev. Mod. Phys.   \textbf{80}, 3, 1161-1193 (2008).
 
 \bibitem{12}
S. F. Radford and W. W. Repko, Potential model calculations and predictions for heavy quarkonium,  Phys. Rev. D  \textbf{75} 074031 (2007). 

\bibitem{13}
A. Martin, A Fit of Upsilon and Charmonium Spectra,  Phys. Lett. B.  \textbf{93}, 338 (1980).

\bibitem{14}
C. Quigg and J. L. Rosner, Quarkonium level spacings,  Phys. Lett. B.  \textbf{71}, 153-157 (1977).

\bibitem{15}
A. Martin, A Fit of Upsilon and Charmonium Spectra, Phys. Lett. B.  \textbf{93}, 338 (1980).

\bibitem{16}
H. Grosse, A. Martin, Particle Physics and the  Schr\"{o}dinger equation, Cambridge Monographs on Particle Physics, Nuclear Physics and Cosmology (1997).

\bibitem{17}
 K. B. Vijaya Kumar Bhaghyesh and   A. P. Monteiro, Heavy quarkonium spectra and its decays in a nonrelativistic model with Hulthen potential,  J. Phys. G: Nucl. Part. Phys. \textbf{38} 085001 (2011).
 
 \bibitem{18}
 O. Lakhina and E. S. Swanson, Dynamics properties of charmonium,  Phys. Rev. D \textbf{74} 014012 (2006). 
 
 \bibitem{19}
 K. A. Olive et al. [Particle Data Group Collaboration], Review of Particle Physics,   Chin. Phys. C \textbf{38} 090001 (2014).
 
 \bibitem{20} 
 B. Q. Li and K. T. Chao, Higher Charmonia and X,Y,Z states with Screened Potential, Phys. Rev. D \textbf{79} 094004 (2009). 
 
 \bibitem{21}
 T. Barnes, S. Godfrey and E. S. Swanson, Higher charmonia, Phys. Rev. D \textbf{72}, 054026 (2005).
 
 \bibitem{22}
  W.-J. Deng, H. Liu, L.-C. Gui, X.-H. Zhong, Charmonium spectrum and electromagnetic transitions with higher multipole contributions, Phys. Rev. D \textbf{95} 034026 (2017).

\bibitem{23}
C. Patrignani et al. (Particle Data Group), Chin. Phys. C, \textbf{40}, 100001 (2016) and 2017 update. 

\bibitem{24}
 B. Q. Li and K. T. Chao,  Bottomonium Spectrum with Screened Potential, Commun. Theor. Phys. \textbf{52}  653-661 (2009). 

\bibitem{25}
W.-J. Deng, H. Liu, L.-C. Gui, X.-H. Zhong,  Spectrum and electromagnetic transitions of bottomonium, Phys. Rev. D \textbf{95} 074002 (2017).

 \bibitem{26}
S. Godfrey and K. Moats, Bottomonium Mesons and Strategies for their Observation, Phys. Rev. D \textbf{92}, 054034 (2015).

\bibitem{27}
J. Segovia, P. G. Ortega, D. R. Entem and F. Fernandez, Bottomonium spectrum revisited, Phys. Rev. D \textbf{93}, 074027 (2016).

\bibitem{28}
R. Van Royen and V. F. Weisskopf, Nuovo Cimento A \textbf{50}, 617 (1967)

\bibitem{29}
E. Braaten and S. Fleming, QCD radiative corrections to the leptonic decay rate of the Bc meson, Phys. Rev. D \textbf{52} 181, (1995).

\bibitem{30}
W. Kwong, P. B. Mackenzie, R. Rosenfeld and J. L. Rosner, Quarkonium annihilation rates, Phys. Rev. D. \textbf{37} 3210, (1988).

\bibitem{31}
J.T. Laverty, S. F. Radford, W. W. Repko, arXiv:hep-ph/0901.3917v3, (2009).

\bibitem{32}
J. Segovia, D. R. Entem, F. Fernandez, E. Hernandez, Constituent Quark Model Description of Charmonium Phenomenology, Int. J. Mod. Phys. E \textbf{22} 10, 1330026, (2013).

\bibitem{33}
W. Lucha, F. Schoberl, Effective potential models for hadrons HEPHY-UB 62P1/95 UW THPH 1995-16.

\bibitem{34}

L. Vengyel, Yu. Fekete, I. Haysak, A. Shpenik,  Calculation of hyperfine splitting in mesons using configuration interaction approach, Eur. Phys. J. C \textbf{21}(2) 355-359, (2001).

\bibitem{35}

C. Quigg, J. L. Rosner, Quarkonium level spacings, Phys. Lett. B \textbf{77}(1) 153-157 (1977).


\bibitem{36}

J. L. Richardson, The heavy quark potential and the $\Upsilon, J/\psi$ systems,  Phys. Lett. B \textbf{82}(2) 272-274 (1979).

\bibitem{37}

W. Buchmüller and S. -H. H. Tye, Quarkonia and quantum chromodynamics, Phys. Rev. D \textbf{24} 132 (1981).

\bibitem{38}

E. Eichten and F. Feinberg, Spin-dependent forces in quantum chromodynamics, Phys. Rev. D \textbf{23} 2724 (1981).











\end{thebibliography}
\end{document}